\documentclass[iop]{emulateapj}
\usepackage[colorlinks=true,citecolor=blue]{hyperref}

\newcommand{\mcolor}[2]{\mbox{[${#1}$]--[${#2}$]}}
\newcommand{\nh}{\mbox{$n_{\mathrm{H}}$}}
\newcommand{\rmxaa}{Rev. Mexicana Astron. Astrofis.} %
\newcommand{\bettertfrac}[2]{ 
    \dimen1 = \the\fontdimen6\font
    \dimen2 = 0.8\dimen1
    \dimen3 = \the\fontdimen5\font
    \dimen4 = 0.1\dimen3
    \begingroup
    \fontsize{\the\dimen2}{\the\fontdimen6\font}\selectfont
    \raisebox{\dimen4}{${#1}$}/\raisebox{-\dimen4}{$\!{#2}$}
    \endgroup
    }
\newcommand{\mtfrac}[2]{
    \dimen1 = \the\fontdimen5\font
    \dimen2 = 0.3\dimen1
    \dimen3 = -0.2\dimen1
    \mbox{\raisebox{\dimen2}{${#1}$}/\raisebox{\dimen3}{${#2}$} 
    }
    }

\slugcomment{Accepted for publication in the Astrophysical Journal Letters}

\shorttitle{Spitzer IRAC Color Diagnostics}
\shortauthors{Ybarra et al.}

\begin{document}

\title{Spitzer IRAC Color Diagnostics for Extended Emission in Star Forming Regions}

\author{Jason E. Ybarra\altaffilmark{1}}
\email{jybarra@astro.unam.mx}

\author{Mauricio Tapia\altaffilmark{1}}
\author{Carlos G. Rom{\'a}n-Z{\'u}{\~n}iga\altaffilmark{1}}

\author{Elizabeth A. Lada\altaffilmark{2}}

\altaffiltext{1}{Instituto de Astronom{\'i}a, 
Universidad Nacional Aut{\'o}noma de Mex{\'i}co, 
Unidad Acad{\'e}mica en Ensenada,
Km 103 Carr. Tijuana-Ensenada,
22860 Ensenada BC, Mex{\'i}co}
\altaffiltext{2}{Department of Astronomy, University of Florida,
211 Bryant Space Science Center,
Gainesville, FL 32611, USA}

\begin{abstract}
The infrared data from the {\it Spitzer Space Telescope} has provided 
an invaluable tool
for identifying physical processes in star formation.
In this study we calculate the IRAC color space of UV fluorescent H$_{2}$ and 
Polycyclic Aromatic Hydrocarbon (PAH) 
emission in photodissociation regions (PDRs)
using the Cloudy code %
with PAH opacities from \citet{DraineLi2007}.
We create a set of color diagnostics that can be applied
to study the structure of PDRs and to
distinguish between FUV excited
and shock excited H$_2$ emission.
To test this method
we apply these diagnostics to {\it Spitzer} IRAC data of NGC 2316.
Our analysis of the structure of the PDR is consistent with previous 
studies of the region. In addition to UV excited emission, we identify shocked gas 
that may be part of an outflow originating from the cluster.
\end{abstract}


\section{Introduction}

The infrared data from the {\it Spitzer Space Telescope} has provided invaluable insights 
in the field of star formation.
In addition to emission from young stellar objects (YSOs), 
non stellar extended emission is present 
in the {\it Spitzer} images of star forming regions. 
Mid-infrared bubbles trace regions of high mass star formation 
where bright shells of Polycyclic Aromatic Hydrocarbon (PAH) emission surround areas 
filled with 24 $\mu$m dust emission 
\citep{Simpson2012}.
In other cases absorption of Galactic background emission reveals Infrared Dark Clouds (IRDCs) 
where star formation is in its earliest stages \citep{Rathborne2006,Butler2009}.
When protostellar outflows are present,
the emission from the associated shocked gas is detected in all 
4 {\it Spitzer} Infrared Array Camera (IRAC) bands, 
primarily due to a multitude of ro-vibrational molecular hydrogen (H$_2$) emission lines 
\citep{Noriega-Crespo2004,Velusamy2014}. 
In IRAC imaging of high mass star forming regions
the presence of extended 4.5 $\mu$m emission 
has been used to identify massive embedded YSOs \citep{Cygnowski2009}.

\citet{YbarraLada2009} calculated the IRAC color space
for shocked H$_2$ emission excited through 
H--H$_2$, He--H$_2$, and H$_2$--H$_2$ collisions. 
They found the location of shocked H$_2$ in 
\mcolor{3.6}{4.5} vs. \mcolor{4.5}{5.8}
color space
to be a function of gas temperature and volume density. 
The calculated color space for high temperature shocked gas 
was found to be consistent with the empirical color cut
used by
\citet{Gutermuth2008}
to distinguish between YSOs and shocked emission. 
\citet{Ybarra2010} 
included emission from the CO $\nu$=1--0 band in their calculations 
of the color space of shocked gas.
Analysis using this color space can be used to systematically search for outflows and
to study their structures and interactions with environments \citep{Ybarra2010, Giannini2013}.

While H$_2$ emission is a good tracer of shocked gas from outflows, emission from 
H$_2$ can also arise from UV fluorescence in cold gas \citep{BlackDalgarno1976,BlackvanDishoeck1987}.
This often occurs in photo-dissociation regions (PDRs) surrounding young high mass stars.
UV radiation can have an effect on star formation; it can either help disperse 
the remnant gas ending the formation of stars in the region or trigger a subsequent generation 
of star formation. 
UV fluorescent H$_2$ emission is due to the absorption of far-ultraviolet (FUV) photons 
by H$_2$ into its excited electronic states.
These excited molecules will either decay into the ground electronic state continuum and 
disassociate or decay into discrete ro-vibrational levels of the ground electronic state 
and then cascade 
through ro-vibrational transitions giving rise to infrared photons \citep{BlackDalgarno1976}. 
The resulting line ratios from the H$_2$ fluorescent emission can often mimic 
those of shock heated gas. 
Within PDRs small PAH grains also absorb FUV photons and re-emit in the infrared. 
This PAH emission is often found to be correlated with H$_2$ emission in H$_2$ shells 
surrounding PDRs \citep{Velusamy2008}.

\cite{DraineLi2007} calculated the IRAC fluxes for PAH emission varying radiation 
strength and grain size distribution. 
In this paper we extend this analysis to include emission
from H$_2$ fluorescent emission.
We calculate the IRAC color space for PDR regions
considering geometry of the gas cloud and
varying the ratio of incident FUV radiation to gas density.
We create a set of color diagnostics that can be applied
to study the structure of PDRs. Additionally,
we find that color analysis can 
be used to distinguish between
FUV excited and shock excited H$_2$ emission.
Finally, we apply these diagnostics to {\it Spitzer} IRAC data of NGC 2316 as an example
of the usefulness of this method.

\section{Calculations}

PDR calculations were performed with version 13.02 of Cloudy, last described by \citet{Ferland2013}.
This code includes a sophisticated model of the H$_2$ molecule described by \citet{Shaw2005}. 
The code self-consistently calculates excitation, formation, dissociation, 
and ortho-para conversion for H$_2$.
The calculations of the ro-vibrational level populations are performed by
following the photo-excitation of H$_2$ into excited electronic states
and subsequent decay into ground state ro-vibrational levels.
The code includes the ability to use
PAH absorption cross sections from \citet{DraineLi2007}.
We used the grain size distribution of \citet{WeingartnerDraine2001a}
from 3.5 \AA\ to 1 $\mu$m.
We considered the total C abundance in the log-normal parts of the size distribution 
to be $b_c = 5.52 \times 10^{-5}$.
The charge state of a PAH grain affects its opacity. \citet{DraineLi2007}
provide cross sections for both neutral grains and charged grains.
The current version of the Cloudy code is unable to 
self-consistently determine the opacity of the grains from their
charge state. In order to work around this limitation we run the code using
an iterative process. 
First, an initial run of the Cloudy code solved the grain ionization-recombination
balance equations \citep{vanHoof2004} considering a fully neutral PAH distribution
in order to obtain ionization fraction distributions as a function of
grain size and depth into the cloud.
Second, the gas cloud is divided into zones where each zone will have its own
grain ionization distribution. For each zone, two 
separate grain size distributions were then created; one for neutral PAHs and one for ionized PAHs. 
The code is then run again on each zone where the input radiation of that zone was the output radiation of its 
preceding zone. 

We considered a simple layout and geometry of a gas cloud to investigate the color space. 
The volume density of the gas was kept constant and 
the shape of the external FUV radiation field was set to the Draine field \citep{Draine1978}.
We ran a grid of models varying the ratio of incident FUV flux to gas density
($-2.0 \leq \log(G/\nh) \leq 0.5$)\footnote{The FUV flux $G$ is in units of $1.6 \times 10^{-3}$ erg cm$^{-2}$ s$^{-1}$}
through the range of gas density, \nh = $10^{3}$--$10^{5}$ cm$^{-3}$.
Using obtained emissivities and opacities for each zone, we applied the
radiative transfer equation to calculate the spectra at various lines of sight through the gas.
We also investigate different gas geometries by placing slabs of constant thickness at 
various depths of $A_{V}$ = \{0.1, 1.0, 2.0, 3.0, 4.0, 5.0\} mag
both in front of and behind the incident radiation.

\section{Color Space Diagnostics}

The resulting infrared spectra 
from our models
were converted into IRAC colors
using IRAC spectral response and calibration data \citep{Reach2005,Hora2008}.

\subsection{IRAC color space for PDRs}

\begin{figure}
 \plotone{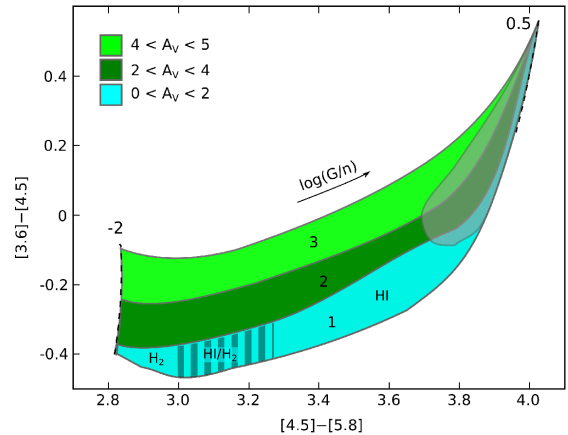}
 \caption{IRAC \mcolor{3.6}{4.5} vs. \mcolor{4.5}{5.8} color-color plot indicating
 the region occupied by PDRs (PAH grains and UV fluorescent H$_2$).
 Region 1 of color-color space is occupied by
 gas clouds of thickness $A_{V} \sim$ 0--2 mag.
 Region 2 of color-color space is occupied by either
 a gas cloud of thickness $A_{V} \sim$ 2--4 mag in front of the
 incident UV source or a gas cloud ($A_{V} > 2$) behind the incident UV source.
 Region 3 indicate a cloud thickness $A_{V} \sim$ 4--5 mag
 in front of the incident FUV source.
 The grey region at \mcolor{4.5}{5.8} $>$ 3.6 indicates region of color space
 with significant degeneracy.
 }
\end{figure}

There is often degeneracy for physical parameters of emitting material in color space, thus the 
usefulness of color space analysis in extracting information is 
restricted to regions in color space that reveals breaks in degeneracy.
From our investigation of the calculated IRAC colors, we find
that the \mcolor{3.6}{4.5} vs. \mcolor{4.5}{5.8} color space 
in the region where 2.9 $<$ \mcolor{4.5}{5.8} $<$ 3.6 
has minimal degeneracy and is useful as a tool 
for analysis. We divide this region
into 3 sub-regions. Figure 1 shows the IRAC \mcolor{3.6}{4.5} vs. \mcolor{4.5}{5.8} color-color plot 
indicating these various regions.

{\emph{Region 1}}.---
This region of color space is defined by
a lower limit
\begin{eqnarray}
 \mcolor{3.6}{4.5} &> 0.15(\mcolor{4.5}{5.8}) - 0.92 \nonumber\\
 \mbox{for } &3.0 < \mcolor{4.5}{5.8} < 3.2 \nonumber\\
 \mcolor{3.6}{4.5} &> 0.35(\mcolor{4.5}{5.8}) - 1.56 \nonumber\\
 \mbox{for } &3.2 < \mcolor{4.5}{5.8} < 3.6 \nonumber
\end{eqnarray}
and upper limit
\begin{eqnarray}
 \mcolor{3.6}{4.5} &< 0.27(\mcolor{4.5}{5.8}) - 1.19 \nonumber\\
 \mbox{for } &3.0 < \mcolor{4.5}{5.8} < 3.3 \nonumber\\
 \mcolor{3.6}{4.5} &< 0.50(\mcolor{4.5}{5.8}) - 1.95 \nonumber\\
 \mbox{for } &3.3 < \mcolor{4.5}{5.8} < 3.6. \nonumber
\end{eqnarray}
This region of color space is mostly 
occupied by thin PDRs of thickness $A_{V} \sim$ 0 to 2 mag.
The colors from a plane-parallel PDR model would be
found in this region of color space.

{\emph{Region 2}}.--- 
This region of color space is defined by
an upper limit of
\begin{eqnarray}
 \mcolor{3.6}{4.5} &< 0.35(\mcolor{4.5}{5.8}) - 1.31 \nonumber\\
 \mbox{for } &3.0 < \mcolor{4.5}{5.8} < 3.6. \nonumber
\end{eqnarray}
This region of color space is occupied by two
cases of PDRs: a) A gas cloud of thickness $A_{V} \sim \mbox{2--4}$ mag
in front of the incident FUV source, or b) A thick gas cloud 
behind the incident FUV source. For a thick cloud behind the FUV source, material
beyond $A_{V} > 2$ mag does not significantly contribute to the total line of sight
emission. Additionally the line of sight emission will begin to include 
a contribution from fluorescent H$_2$ lines.

{\emph{Region 3}}.---
This region of color space is defined by
an upper limit of
\begin{eqnarray}
 \mcolor{3.6}{4.5} &< 0.20(\mcolor{4.5}{5.8}) - 0.72 \nonumber\\
 \mbox{for } &3.0 < \mcolor{4.5}{5.8} < 3.4 \nonumber\\
 \mcolor{3.6}{4.5} &< 0.70(\mcolor{4.5}{5.8}) - 2.42 \nonumber\\
 \mbox{for } &3.4 < \mcolor{4.5}{5.8} < 3.6. \nonumber
\end{eqnarray}
Incident UV source is behind a gas cloud of thickness $A_{V} \sim \mbox{4--5}$. 
The increasing \mcolor{3.6}{4.5} color partly due to the increasing  
contribution of UV fluorescent H$_2$ emission in the line of sight spectra.

Another result of our calculation is that the \mcolor{4.5}{5.8} color
is an approximate indicator of the $G/n_{\rm{H}}$ ratio where the line of sight intersects
the plane of the incident FUV. 
We fit the following analytic form to this relationship:
\begin{displaymath}
\log(G/n_{\rm{H}}) = -19.7 + 10.0a - 1.35a^{2},
\end{displaymath}
where $a = \mcolor{4.5}{5.8}$ and $2.8 < a < 3.8$.
This relationship is due in part to the dependence of the 
ionization fraction on $G/n_{\rm{H}}$. Larger $G/n_{\rm{H}}$
results in larger overall ionization fraction of the PAHs and thus
the increase PAH feature (5.270, 5.700, 6.220 $\mu$m) emission in the 5.8 $\mu$m band
relative to the continuum emission in the 4.5 $\mu$m band. 
Our simple model does not take into account PAH destruction from
strong FUV fields which could effect this color relationship.
At $G/\nh \lesssim 0.04$, H$_2$ begins forming due to self-shielding, and thus the 4.5 $\mu$m band
will also include emission from UV fluorescent H$_2$ lines \citep{Hollenbach1999}.

Finally, we find that the \mcolor{4.5}{5.8} color can be used to probe the
phase of the gas (molecular, mixed, or atomic)
where the line of sight intersects the plane of the incident FUV.
The emission of gas within the \ion{H}{1}/H$_2$ transition zone will have a
\mcolor{4.5}{5.8} color within the range 3.0--3.3. 
Mostly molecular gas ($n({\rm{H_{2}}})/n_{\rm{H}} > 0.3$)
will have \mcolor{4.5}{5.8} $<$ 3.0 
and mostly atomic gas ($n({\rm{H_{2}}})/n_{\rm{H}} < 0.25$)
will have \mcolor{4.5}{5.8} $>$ 3.3 (see Fig. 1).
Our calculations show that within the \ion{H}{1}/H$_{2}$ transition zone the
contribution of H$_2$ ro-vibrational lines to total emission is $\sim$ 5-10\% in
the 3.6 $\mu$m band and $\sim$ 10-20\% in the 4.5 $\mu$m band.
Our model did not include the possible freeze out of small
PAH molecules in cold molecular gas ($\log(G/\nh) \lesssim -2$)
which would reduce the contribution of PAH emission in the IRAC bands.

It should be noted that external foreground extinction can affect the IRAC colors, 
although this effect is not very strong due to the relative flatness of the MIR 
extinction curve \citep{Lutz1996}. 
Application of most MIR extinction laws result in the \mcolor{4.5}{5.8} color to be negligibly 
effected by extinction \citep{Lutz1999, Indebetouw2005}. 
The \mcolor{3.6}{4.5} color appears to have a stronger dependence on extinction, 
however, an $A_{V}$ = 10 mag extinction only results in an increase in the 
\mcolor{3.6}{4.5} color of $\sim$ 0.1 \citep{Indebetouw2005, Chapman2009}. 
However, due to variations of the MIR extinction law with environment, 
the effect of extinction on the \mcolor{3.6}{4.5} color may be even less \citep{SWang2013}. 

\subsection{Distinguishing between shocked and UV fluorescent molecular hydrogen}

It is often the case that the excitation mechanism for H$_2$ emission is unknown,
and in some cases is necessary to understand the nature of the emission.
For example the H$_2$ emission from externally illuminated structures
such as pillars or globulettes can have a similar morphology 
to bow shocks from protostellar outflows. 
We propose the following simple one-color diagnostic that should suffice in most cases of interest
where H$_2$ has already been detected (e.g., NIR H$_2$ 2.12 $\mu$m imaging):
\begin{eqnarray}
 \mbox{Shocked:} &\mcolor{3.6}{4.5} \geq 0.5\nonumber\\
 \mbox{UV Flourescent:} &\mcolor{3.6}{4.5} < 0.5\nonumber
\end{eqnarray}
This diagnostic is particularly useful when only data from the first
two IRAC channels are available, such as 
data from the {\it Spitzer} Warm Mission.  
For analysis including the 5.8 $\mu$m band, the region of color space
defined for shocked H$_2$ emission is
\begin{eqnarray}
 \mcolor{3.6}{4.5} &> -0.24\mcolor{4.5}{5.8} + 1.26 \nonumber\\
 \mbox{for } &0.25 < \mcolor{4.5}{5.8} < 1.5\nonumber\\
 \mcolor{3.6}{4.5} &> 0.8\mcolor{4.5}{5.8} + 1.0\nonumber\\
 \mbox{for } &-0.5 < \mcolor{4.5}{5.8} \leq 0.25\nonumber
\end{eqnarray}
\citep{Ybarra2010}, whereas, gas containing UV fluorescent H$_2$ will 
be in the PDR color space defined in section 3.1.
Figure 2 shows a larger color-color map showing the location of shocked H$_2$  
emission in relation to UV excited PDR emission.
\begin{figure}
 \plotone{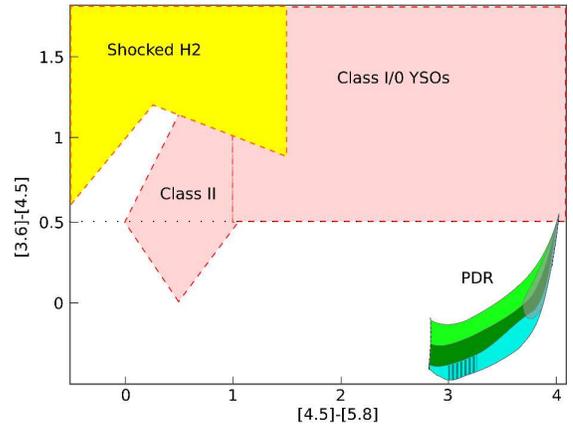}
 \caption{{\it Spitzer} IRAC \mcolor{3.6}{4.5} vs. \mcolor{4.5}{5.8} color-color plot
 showing the location of shocked H$_2$ emission in relation to UV excited PDR emission.}
\end{figure}

\subsection{Background Estimation Method}

In order to measure the emission in each pixel it is necessary to 
model the background. 
For small scale structures, such as shocked H$_2$ knots against a variable 
background, a ring filter can be used to effectively model the background
\citep{YbarraLada2009}.
However, for the analysis of more extended emission, such as that from PDRs, 
we use a
Bayesian intensity estimator developed by \citet{Bijaoui1980}.
The pixel intensity histogram is modeled with a Gaussian
distribution for noise centered near the sky background level and 
an exponential distribution prior to describe the asymmetry of the histogram
due to non-background emission.
The model has the following analytic form:
\begin{displaymath}
 p(I) = \frac{1}{2a}\exp(\mtfrac{\sigma^{2}\!}{2a^{2}})\exp[-(I-s)/a]\mbox{erfc}\left(\frac{\alpha}{\sqrt{2}}\right),
\end{displaymath}
where 
\begin{displaymath}
 \alpha = \frac{\sigma}{a} - \frac{(I-s)}{\sigma},
\end{displaymath}
$I$ is the measured pixel intensity, $s$ is the background level,
$\sigma$ is the width of the Gaussian, and $a$ is the scale 
parameter of the exponential distribution\footnote{erfc is the complementary error function as defined 
by $\mbox{erfc}(z) = (\bettertfrac{2}{\sqrt{\pi}}) \int_{z}^{\infty} \exp(-t^{2}) dt$}. 
We fit the function $p(I)$ to the intensity histogram to 
obtain the values $a$, $\sigma$, and $s$. 
The Bayesian estimator of the true intensity is
\begin{displaymath}
 \hat{i} = I - s - \frac{\sigma^{2}}{a} + \sqrt{\frac{2}{\pi}}
 \frac{\sigma e^{-\alpha^{2}/2}}{\mbox{erfc}(\mtfrac{\alpha}{\!\sqrt{2}})}.
\end{displaymath}
Therefore for every pixel in the image we are able to estimate the intensity 
above the background, and subsequently with the registered images we are able to
calculate the color for each pixel.

\section{Application of the Method: NGC 2316}

The usefulness of any method needs to be tested. 
We apply this method to archival {\it Spitzer} IRAC images of the 
NGC 2316 (=Parsamyan 18 =L1654, ($\alpha$,$\delta$)(J2000) = 
$6^{\mathrm h}59^{\mathrm m}40^{\mathrm s}$,
$-7\arcdeg46\arcmin36\arcsec$)
from program \dataset[ADS/Sa.Spitzer#10707712]{3290 (PI: Langer)}. 
This region is a site of star formation containing a young (2--3 Myr) embedded cluster
with line of sight extinction $\langle A_{V}\rangle$ = 4.5 mag
at a distance of 1.1 kpc\citep{Teixeira2004}. 
Within this region exists a PDR generated by the FUV radiation from a 
ZAMS B3 star \citep{Lopez1988,Ryder1998}.
\notetoeditor{This figure would look best printed across both columns} 
\begin{figure*}
 \plottwo{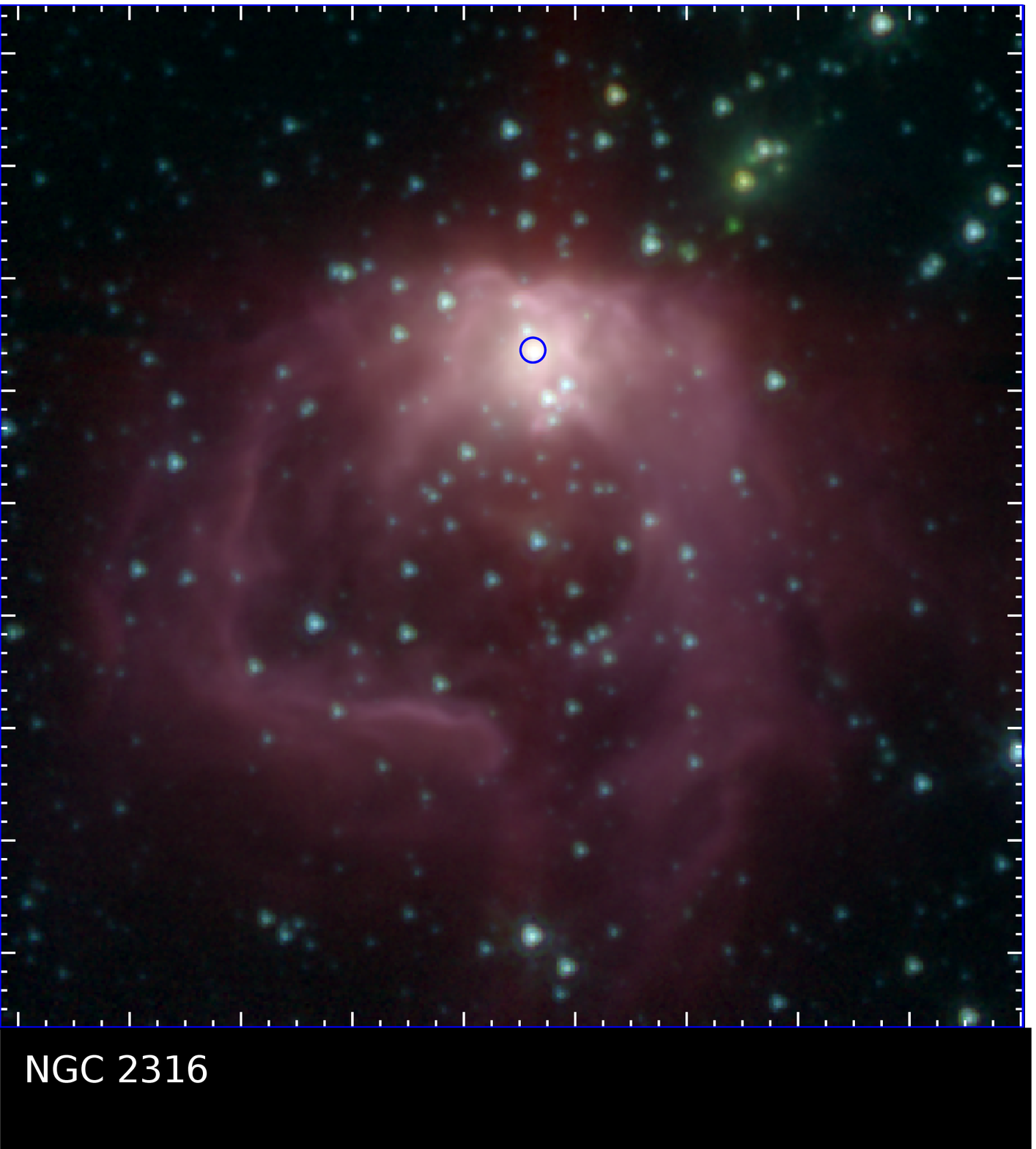}{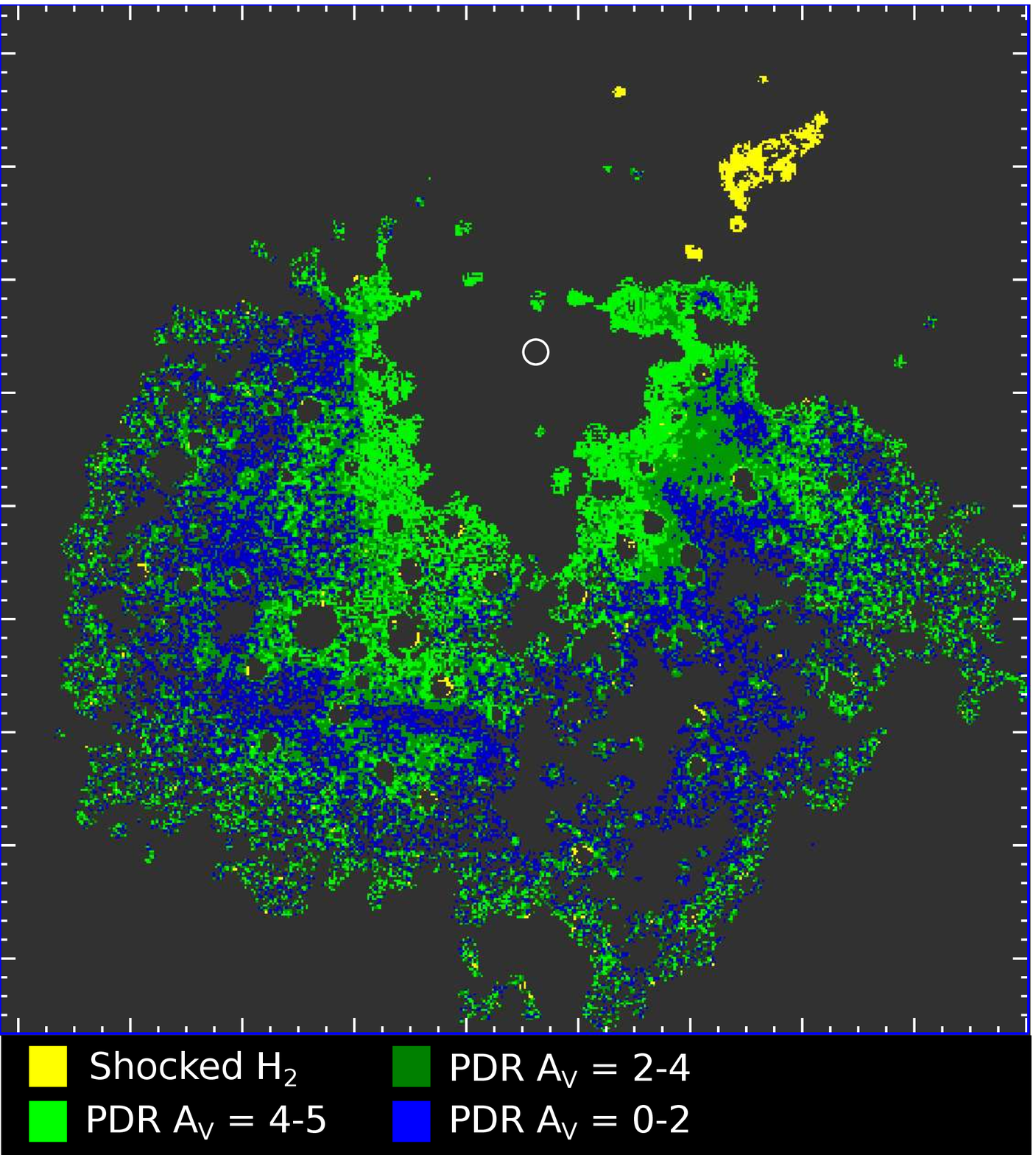}
 \caption{a) {\it Spitzer} IRAC 3-color (R: 5.8 $\mu$m, G: 4.5 $\mu$m, B: 3.6 $\mu$m) 
 image of NGC 2316. Image is centered at ($\alpha$,$\delta$)(J2000) = 
 (6$^{\mathrm h}$59$^{\mathrm m}$42\fs0, $-$7\arcdeg47\arcmin11\arcsec) 
 with a FOV of 4.54\arcmin $\times$ 4.54\arcmin.
 The small circle indicates the location of the B3 star.
 b) Map of identified emission from color analysis. 
 Yellow indicate pixels whose IRAC colors are consistent with that of shocked H$_2$ emission.
 Blue, dark green, and light green indicate pixels with IRAC colors consistent 
 with PDRs with line of sight thicknesses of $A_{V} \sim$  
 0--2 mag, 2--4 mag, and 4--5 mag, respectively.
 }
\end{figure*}
Figure 3a shows a {\it Spitzer} IRAC 3-color image of the cluster. 
Figure 3b shows a map of the results from our color analysis.
The color-coded yellow regions indicate pixels whose
IRAC colors are consistent with that of shocked H$_2$ emission.
Blue, dark green, and light green indicate pixels with IRAC colors consistent 
with PDRs with line of sight thicknesses of $A_{V} \sim$  
0--2 mag, 2--4 mag, and 4--5 mag, respectively.

We find the emission around the B3 star to have colors consistent with PDRs with 
line of sight thickness of $A_{V} > 4$ mag. 
We are able deduce that this PDR results from an exciting source embedded within the cloud.
which is consistent with previous studies of the region \citep{Ryder1998,Velusamy2008}
Our analysis is also consistent with the results from the NIR H$_2$ study
by \citet{Ryder1998}. We identify five of the eight H$_2$ emission regions in our color
analysis (R2=North Peak, R4=S Peak, R6=NW Arc, R7=W Arc, R8=P 18 NW). 
\begin{figure}
 \plotone{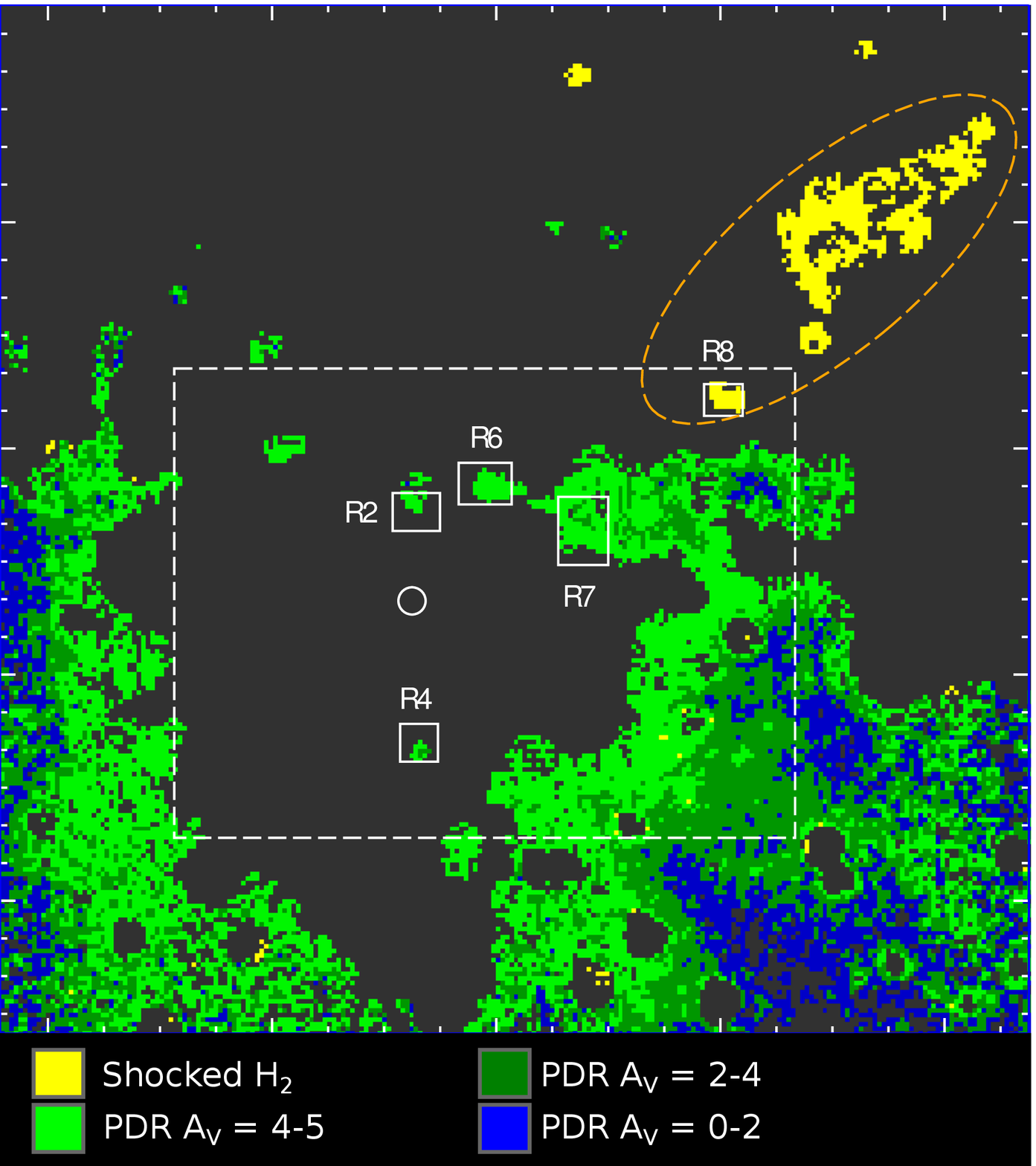}
 \caption{Map of identified emission from color analysis.  
 The solid boxes indicate the locations of five of the 
 H$_2$ emission regions studied by \citet{Ryder1998}.
 The dashed box indicates FOV of the \citet{Ryder1998} H$_2$ study.
 The orange ellipse indicates the proposed outflow.
 The circle represents the location of the B3 star.
 The map is centered at ($\alpha$,$\delta$)(J2000) = 
 (6$^{\mathrm h}$59$^{\mathrm m}$40\fs7, $-$7\arcdeg46\arcmin17\arcsec)
 with a FOV of 2.27$^{\prime}$ $\times$ 2.27$^{\prime}$.} 
\end{figure}
Figure 4 shows a color analysis map with the locations of the five emission regions 
indicated by solid boxes. Four of these regions (R2, R4, R6, R7) were found by
\citet{Ryder1998} to have H$_2$ emission excited through UV fluorescence. 
These four regions are part of the H$_2$ shell observed by \citet{Velusamy2008}. 
Our color analysis of these four regions reveals they have colors consistent
with PDR emission from a cloud with thickness $A_{V} \sim$ 4--5 mag.
Additionally, those four regions have $\mcolor{3.6}{4.5} \sim$ 3.2--3.3, 
corresponding to the \ion{H}{1}/H$_2$ transition zone
where H$_2$ is expected to strongly emit \citep{Hollenbach1999}.

\citet{Ryder1998} found that the region R8 has a H$_2$ (1--0)/(2--1) S(1) line ratio consistent
with shock excitation and therefore suggest R8 may be part of an outflow
possibly associated with the previously observed CO outflow found in the region \citep{Fukui1993}. 
Our analysis reveals this region to have IRAC colors consistent with shock excited H$_2$ and
therefore confirming the previous result. Northwest of R8 our color
analysis reveals more emission that may be shock excited
This emission has an elongated structure
that along with R8 may be part of an outflow originating from the B3 star or a nearby 
neighbor. 
The projected length of this structure is 0.27 pc (51\arcsec) and the projected 
distance from the B3 star to the furthest edge is 0.53 pc (99\arcsec).

\section{Conclusions}
We created a set of {\it Spitzer} IRAC color diagnostics 
to study the extended MIR emission
observed in star forming regions. 
For PDRs, 
color diagnostics allow one to probe
the depth of the region, the phase of the gas,
and estimate the ratio of FUV to gas density.
Additionally, we show that color analysis can distinguish between FUV excited
and shock excited  H$_2$ emission.
We present a one-color diagnostic that is able to make use 
of new data from the {\it Spitzer} Warm Mission
for distinguishing the excitation processes. 
To demonstrate the usefulness of our method
we applied it to {\it Spitzer} IRAC observations
of NGC 2316 and find our results are consistent with previous studies
of the nebula. 
Additionally we identify shocked gas which may be part of an outflow
originating from the cluster.

\acknowledgments

We thank the referee for suggestions in improving this manuscript.
JY thanks Grace Wolf-Chase, Kim Arvidsson and Charles Kerton for useful discussions
during the preparation of this manuscript, and Peter Barnes
for discussions inspiring this study.
JY acknowledges support from a UNAM DGAPA postdoctoral fellowship.
Partial support from PAPPIT-IN101813 is acknowledged.
CRZ and JY acknowledge additional support from
program CONACYT CB2010 152160 Mexico.
This work is based in part on archival data obtained with
the {\it Spitzer Space Telescope}, which is operated by the Jet
Propulsion Laboratory, California Institute of Technology under
a contract with NASA.

{\it Facilities:} \facility{Spitzer(IRAC)}.

\clearpage

\end{document}